\documentclass[12pt]{iopart}

\usepackage{amssymb}
\usepackage{amsbsy}
\usepackage{graphicx}
\usepackage{bbm}
\usepackage{bbold}
\usepackage{hyperref}

\newcommand{\bise}{Bi$_2$Se$_3\,$}

\usepackage{fancyhdr}
\begin{document}

\title{Band bending at the surface of \bise studied from first principles}

\author{P. Rakyta$^{1,2}$,  B. Ujfalussy$^{3}$, and L. Szunyogh$^{1,2}$}

\address{$^1$ Department of Theoretical Physics, Budapest University of Technology and Economics, H-1111 Budafoki \'ut. 8, Hungary} 
\address{$^2$ MTA-BME Condensed Matter Research Group, Budapest University of Technology and Economics, H-1111 Budafoki \'ut. 8, Hungary}
\address{$^3$ MTA Wigner RCP, H-1525 Budapest, PO. Box 49}

\ead{rakytap@bolyai.elte.hu}

\begin{abstract}
The band bending (BB) effect on the surface of the second-generation topological insulators implies a serious challenge to design transport devices.
The BB is triggered by the effective electric field generated by charged impurities close to the surface and by the inhomogeneous charge distribution of the occupied surface states. 
Our self-consistent calculations in the Korringa-Kohn-Rostoker framework showed that in contrast to the bulk bands, the spectrum of the surface states is not bent at the surface.
In turn, it is possible to tune the energy level of the Dirac point via the deposited surface dopants.
In addition, the electrostatic modifications induced by the charged impurities on the surface induce long range oscillations in the charge density.
For dopants located beneath the surface, however, these oscillations become highly suppressed.
Our findings are in good agreement with recent experiments, however, our results indicate that the concentration of the surface doping cannot be estimated from the energy shift of the Dirac cone within the scope of the effective continuous model for the protected surface states.
\end{abstract}

\pacs{73.20.At;}
 
 


\section{Introduction}

The theoretical discovery of the second-generation topological insulators \cite{TI_Zhang} (2GTIs) triggered an intensive experimental effort to observe the predicted surface states \cite{TI_SbTe_exp,time_dep_ARPES,bise_arpes2,bise_arpes,SARPES_bite,dichroism,SARPES_bite2,CucontorolledBB,FeonBiSe} (SSs) being protected by time-reversal symmetry \cite{fu}.
It turned out, that the physical properties of the prepared samples are greatly affected by the electron acceptor/donor impurities, that can be found either in the bulk or on the surface.
The created \bise samples are typically electron doped by the inner point defects \cite{BiSe_vacancies_exp,BiSe_vacancies,Cadopants_exp}.
However, it has been shown that the Fermi level can be tuned by the insertion of further bulk dopants into the system \cite{SARPES_bite,bise_arpes,Cadopants_exp}.
The presence of the charged impurities in the system and the inhomogeneous charge distribution close to the surface generates an effective electrostatic field, which can be probed experimentally, for example by second-harmonic generation \cite{potential_probe1,potential_probe2}.

In addition, the evolved electrostatic field induces a band bending (BB) in the bulk band structure close to the surface, which was successfully observed by angle-resolved photomeisson (ARPES) experiments as well \cite{TI_SbTe_exp,time_dep_ARPES,bise_arpes,surfandbulk,Kdopants_exp1,Cadopants_exp}.
The experimental manipulation of the BB field was also accomplished by the insertion of bulk Cu dopants into the \bise matrix \cite{CucontorolledBB}.
Moreover, ARPES experiments also demonstrated the possibility to shift the Dirac cone by applying charge dopants on the \bise surface \cite{SARPES_bite,Cadopants_exp}.
An experimental evidence for a large shift of the Dirac cone towards the conduction band was also reported by gated terahertz cyclotron resonance measurements performed on thin \bise film \cite{Dirac_shift_exp1}.

Besides these comprehensive experimental studies, numerous theoretical works were also devoted to the description of the physical properties of the 2GTIs, including first principle calculations \cite{TI_Zhang,GW,GW2,KKR_bite,Kdopants,Kdopants2,Bibilayer}, tight binding \cite{Kdopants,TBmodel,TBcikk} or effective continuous \cite{fu} models.
The structure of the 2GTIs can be described by a sequence of weakly bound quintuple layers (QLs), each consisting of five atomic layers.
The effect of the bulk dopants on the electronic behavior of \bise crystal was also addressed in recent theoretical studies \cite{TBcikk,BiSe_doping,Kdopants,Kdopants2}.
In addition, Galanakis \emph{et. al.} \cite{TBcikk} also suggested that the BB profile and the energy of the Dirac point can be controlled by electrostatic effects.
The importance of the BB has been demonstrated in other 2GTI based nanostructures as well, including topological/normal insulator interfaces \cite{BB_TI_I_exp,Dirac_engeneering,BB_TI_I,spatial_Dirac_shift}.

Motivated by these research findings in this work we study the BB effect and the Dirac cone shift on the surface of 2GTIs theoretically.
In particular, we perform screened Korringa-Kohn-Rostoker (SKKR) calculations to examine the role of the charged dopants at \bise surface. 
The slow dynamics of the band bending process suggests that the charge accumulation at the surface is coupled to a much slower surface lattice relaxation \cite{time_dep_ARPES}.
In this work we do not aim to describe the time dependence of the outlined process, but rather to examine the effect of the surface dopants on the band structure.
In addition, according to Ref. \cite{Kdopants2} the lattice relaxation in the presence of adatoms is expected to be negligible small.
Thus, in our surface calculations we consider a rigid lattice excluding any structural relaxation processes of the surface layers.

The rest of the paper is organized as follows.
In Sec.~\ref{sec:method} we present the details of our numerical approach to study the bulk and surface properties of Bi$_2$Se$_3$.
Than in Sec.~\ref{sec:results} we examine the BB and the effect of the surface dopants on the dispersion of the SSs.
Finally we summarize our work in Sec.~\ref{sec:summary}.


\section{Details of the numerical calculations} \label{sec:method}

In order to describe the electron structure of Bi$_2$Se$_3$, we used the relativistic spin-polarized SKKR method \cite{KKR3}.
The first-principles calculations were performed by density functional theory using the local spin-density approximation and the Ceperley-Alder parametrization of the exchange correlation functional \cite{CA_exchange} within the the atomic-sphere approximation (ASA). 
In our calculations we used an angular momentum cutoff $l_{\rm max}=2$.

\begin{figure}[thb]
\centering
\includegraphics[scale=0.4]{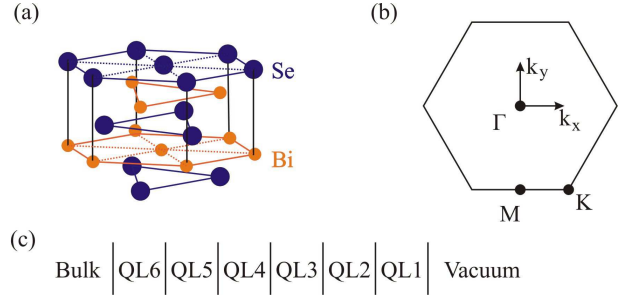}
\caption{(a) The structure of the \bise crystal within one QL.
(b) The two-dimensional Brillouin zone related to the surface terminated by an atomic plane of a QL.
(c) The scheme of the studied system. Six QLs are surrounded by a semi-infinite bulk of \bise crystal from the left and by vacuum from the right.
} \label{fig:lattice}
\end{figure}

The 2GTIs posses a rhombohedral lattice structure were the atoms are located in parallel layers forming a triangular lattice \cite{BiSe_lattice}.
This lattice structure can be described by a periodic sequence of QLs, whereas each of the QL consists of five strongly bound atomic layers.  
The QLs, on the other hand, are weakly bound to each other by van der Waals couplings.
In particular, the QLs in the Bi$_2$Se$_3$ crystal consist of atomic layers Se1-Bi-Se2-Bi-Se1, whereas Se1 and Se2 are selenium atoms at inequivalent geometrical positions.
Fig.~\ref{fig:lattice}(a) shows the structure of one QL in the lattice.
Since the QLs are weakly bound to each other, the crystal surface is favored to be terminated by Se atoms. 
Thus, in our calculations we assumed a flat surface formed by the last Se atomic plane of a QL.
The considered surface has a hexagonal structure, with a 2D lattice constant $a=4.138$ \AA\; and lattice vectors given by $\mathbf{a}_1 = (a, 0,0)$ and $\mathbf{a}_2 = (-\frac{1}{2}a, \frac{\sqrt{3}}{2}a, 0)$, where the axis $z$ is perpendicular to the surface. 
The shortest period in the lattice structure along the $z$ axis is given by three successive QLs \cite{TI_Zhang}.
Still, the periodicity of the Bi$_2$Se$_3$ lattice in the $z$ direction can be well described by a skew lattice vector $\mathbf{a}_3 = (5a, \frac{5}{\sqrt{3}}a, z_{QL})$ (with \mbox{$z_{QL}=9.547$ \AA\;} standing for the hight of one QL) resulting in a period length of one QL only in the direction spanned by $\mathbf{a}_3$. 

According to the SKKR method \cite{KKR3}, we modeled the bulk system by a single QL surrounded by semi-infinite bulk regions from both the left and right sides.
The surface Greens functions of the semi-infinite regions were calculated by an iterative method \cite{iterGreen1,iterGreen2}.
The charge density distribution was calculated by means of the energy-dependent Greens function, integrated up to the Fermi energy.
The position of the Fermi energy was determined to satisfy the total charge neutrality condition.
In particular, because of the ASA used in our calculations, however, the value of the Fermi level is obtained by a numerical error.
Though, the correct position of the Fermi energy is essential for insulator systems.
Several previous works \cite{Lloyd1,Lloyd2,Lloyd3} proposed a procedure to correct the value of the Fermi energy in a self-consistent way based on the Lloyd's formula \cite{Lloyd_orig}.
Following this procedure we obtained the proper Fermi energy by re-normalizing the wave functions in order to obtain the correct space-integrated charge distribution.

For the surface calculations we considered an interface region surrounded by a semi-infinite bulk system from the left and by a vacuum from the right [see Fig.~\ref{fig:lattice}.(c)]. 
The interface region was constructed from six QLs to support a smooth transition of the atomic potentials as we proceed from the bulk layers to the vacuum.
We also included additional (in total eight) vacuum layers between the surface of the interface region and the semi-infinite vacuum side.
From an electrostatic point of view, the atomic potentials (as well as the charge density distribution) were obtained by imposing zero electrostatic field far away from the surface of the crystal.

The described numerical method is sufficient to obtain plausible results for the studied system \cite{KKR_bite}, however, the inclusion of empty spheres between the atomic layers further stabilized our numerical approach.
  \begin{table}[ht]
  \begin{center}
\begin{tabular}{|c|c|c|c|c|}
\hline
Type & x [\AA]& y [\AA]& z [\AA]& $R_{WS}$ [\AA] \\
\hline
\hline
  E1 & 0    & 0    & 0    & 1.447   \\
  Se1 & 2.069   &  1.195    & 1.184    & 1.587   \\
  E2 & 4.138   & 2.389    & 2.1408   & 1.226    \\
  Bi & 6.207   & 3.584    & 2.893    & 1.758    \\
  E3 & 8.276    & 4.778  &  3.664    & 1.1854    \\
  Se2 & 10.3450  &   5.973  &   4.773  &   1.698    \\
  E3 & 12.414  &  7.167   &  5.883   &  1.185    \\
  Bi & 14.483   &  8.362   &  6.654   &  1.758    \\
  E2 & 16.552   &  9.556   &  7.406   &  1.226    \\
  Se1 & 18.621  &  10.751  &   8.363  &   1.587    \\
\hline
\end{tabular}
\caption{The $x$, $y$ and $z$ coordinates and the Wigner-Seitz radii ($R_{WS}$) of the atomic and empty spheres in one QL. Notations Se1 and Se2 [E1, E2, E3] stand for the inequivalent selenium [empty] spheres.
The position of the other spheres in the lattice can be computed via the lattice vectors $\mathbf{a}_1$, $\mathbf{a}_2$ and $\mathbf{a}_3$ (see the text for details). \label{table:spheres}} 
\end{center}
  \end{table}
In our calculations we used identical geometrical parameters for the lattice structure as in Ref. \cite{BiSe_lattice}, however, we optimized the positions and the Wigner-Seitz radii of the empty spheres to reproduce the main features of the experimentally observed band structure of the SSs.\cite{bise_arpes}
Focusing on the direct band gap at the $\Gamma$ point and on the slope of the Dirac cone, the obtained numerical parameters are summarized in Table \ref{table:spheres}.

Finally, the band structure can be obtained from the Bloch spectral function (BSF).
For surface calculations the layer-resolved BSF depends on the energy $E$ and on the parallel momentum $k_{\parallel}$:
\begin{equation}
 A_n(E, \mathbf{k}_{\parallel}) = -\frac{1}{\pi}{\rm Im}{\rm Tr}\int\limits_{\Omega_n} {\rm d}^3r_n\;G^+(\mathbf{r}_n,\mathbf{r}_n,E, \mathbf{k}_{\parallel}), \label{eq:spectral}
\end{equation}
where $G^+(\mathbf{r}_n,\mathbf{r}_n,E, \mathbf{k}_{\parallel})$ is the retarded Greens function at position $\mathbf{r}_n$ in a selected atomic sphere $\Omega_n$ of layer $n$, and the trace is taken over the quantum numbers of the total angular momentum.
Thus, the BSF is ideal to study the surface states, whereas the three-dimensional bulk band structure is projected onto the two-dimensional Brillouin zone (BZ) corresponding to the crystal surface [see Fig.\ref{fig:lattice}(b)].


\section{Band bending and shifting of the Dirac cone by charge dopants} \label{sec:results}

We now turn our attention to the SSs formed on the surface of \bise crystal.
In this section we discuss our results on the BB effect induced by the electric charge accumulation and/or inhomogeneous charge distribution close to the surface.
We also show that the Dirac cone can be shifted in energy due to the effective electric field generated by the deposited surface dopants. 
Our findings are in good agreement with the ARPES experiments \cite{SARPES_bite}.

Making use of the atomic potentials determined by the self-consistent calculations described in Sec. \ref{sec:method} we calculated the layer-resolved BSF in the QLs beneath the surface.
(The band structure calculated for the bulk crystal is presented in appendix \ref{subsec:bulk}.)
\begin{figure}[thb]
\centering
\includegraphics[scale=0.5]{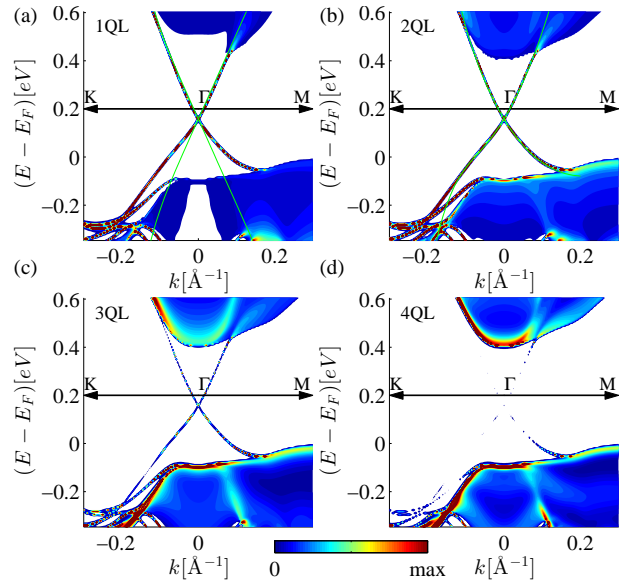}
\caption{The BSF given in Eq.~(\ref{eq:spectral}) and summed over the layers of the (a) 1QL, (b) 2QL, (c) 3QL and (d) 4QL in the interface region plotted along the $K\Gamma M$ cross section of the two-dimensional BZ.
The projection of the bulk bands onto the two-dimensional BZ is shown by a colored areas, while the narrow lines correspond to the bands with low dispersion in the $z$ direction, including the SSs.
The intensity of the SSs rapidly decrease in deeper QLs. 
The results of the SKKR model are compared to the energy bands of the effective model (\ref{eq:TI_spektrum}) using parameter set (a) $v_0\approx3.55$ eV\AA, $\lambda\approx128$ eV\AA$^3$, $1/(2m)=0$, $\alpha=0$ and (b) $v_0^{*}\approx1.63$ eV\AA, $\lambda^{*}\approx108$ eV\AA$^3$, $1/(2m^{*})\approx20$ eV\AA$^3$, $\alpha^{*}\approx78$ eV\AA$^3$ (see the text for details).
The BSF was calculated at complex energies with small imaginary part of $\sim0.7$ meV.
} \label{fig:comparison}
\end{figure}
Fig.~\ref{fig:comparison} shows the calculated BSF summed over the layers of the individual QLs.
The narrow lines correspond to the bands with low dispersion in the $z$ direction, including the SSs.
Within the bulk band gap the dispersion of the protected SSs form a Dirac cone, which is anisotropic in the $k_{\parallel}$ plane.
The Dirac point is located $\sim 250$ meV below the bulk conduction minimum, in good agreement with the ARPES measurements \cite{SARPES_bite}.
The SSs penetrates below the surface up to the third QL, where the intensity of the SSs eventually vanishes. 
The signatures of the bulk band structure, however, can be observed in all the QLs up to the topmost one.
Thus, the SSs spatially overlap with the bulk states, hence they can hybridize.
Consequently, one can observe an increased width of the SS bands in the vicinity of the bulk bands.

The SSs that are not hybridized with the bulk states, on the other hand, can be described by a $2\times2$ effective Hamiltonian proposed by Fu et al.\cite{fu} up to the third order of the momentum $\hbar\mathbf{k} = (\hbar k_x, \hbar k_y)$:
\begin{equation}
 H(\mathbf{k}) = \frac{k^2}{2m} + v_{k}(k_x\sigma_y - k_y\sigma_x) + \frac{\lambda}{2}(k_+^3 + k_-^3)\sigma_z\;. \label{eq:TI_FU_ham}
\end{equation}
Here $v_k = v_0 + \alpha k^2$, $k_{\pm}=k_x\pm{\rm i}k_y$ and $\sigma_{x,y,z}$ stand for the Pauli matrices acting in the spin space. 
Parameters $m$, $v_0$, $\alpha$ and $\lambda$ are to be determined either from first principle calculations or from fits to the experimental data.
The energy eigenvalues of the electron states are given by the expression
\begin{equation}
 E_{\pm}(\mathbf{k}) = \frac{k^2}{2m} \pm \sqrt{v_k^2k^2 + \lambda^2k_x^2(k_x^2 - 3k_y^2)^2}\;, \label{eq:TI_spektrum}
\end{equation}
where $\pm$ labels the bands above/below the Dirac point.
The effective mass $m$ introduces an asymmetry between the upper and lower side of the Dirac cone that is, indeed, significant according to the calculated band structure shown in Fig.~\ref{fig:comparison}.
In Refernce \cite{bise_arpes} the parameters of Eq.~(\ref{eq:TI_spektrum}) were fitted to the experimental ARPES data in the upper side of the Dirac cone.
Focusing on the parameters describing the most pronounced features of the dispersion, namely the Fermi velocity $v_0$ and the parameter $\lambda$ responsible for the hexagonal warping, the obtained numerical values were $v_0\approx3.55$ eV\AA, $\lambda\approx128$ eV\AA$^3$, $1/(2m)=0$ and $\alpha=0$.
Fig.~\ref{fig:comparison}(a) compares the  band structure of the effective model to the SKKR band structure. 
As one can see, the effective model with the given parameters reproduces the calculated band structure of the SSs above the Dirac point well.
However, below the Dirac point the agreement of the effective and SKKR model is weak.
To describe the asymmetry between the upper and lower side of the Dirac cone we propose another set of parameters, namely $v_0^{*}\approx1.63$ eV\AA, $\lambda^{*}\approx108$ eV\AA$^3$, $1/(2m^{*})\approx20$ eV\AA$^3$ and $\alpha^{*}\approx78$ eV\AA$^3$.
Using this parameter set we find that both the upper and lower side of the Dirac cone obtained by Eq.~(\ref{eq:TI_spektrum}) is close to the dispersion of the SSs calculated within the SKKR framework.

Comparing the band structure shown in Fig.~\ref{fig:comparison}.(a) to the ones of Figs.~\ref{fig:comparison}.(b)-(d), one can observe the BB in the topmost QL even without any extra charge dopants added to the system.
Due to the effective electrostatic field induced by the inhomogeneous charge distribution close to the surface, the bulk conduction band is repelled upward by about \mbox{$100$ meV} in the first QL compared to the conduction band minimum in the other QLs  (see Fig.~\ref{fig:comparison}). 
A BB of similar magnitude on the surface of pristine \bise crystal was also reported in other DFT calculations \cite{fregoso}.
We expect that charge dopants deposited on the \bise surface further modifies the electrostatic configuration of the top layers.
Indeed, in Hsieh \emph{et. al.} \cite{SARPES_bite} the surface of the \bise sample was dosed by NO$_2$ molecules and accomplished to shift the Dirac point towards the conduction band.
The presence of charged impurities in the sample, like electron donor Se vacancies generated during the sample preparation process or reactive chemical doping \cite{BiSe_vacancies_exp}, also plays a crucial role in the electrostatic properties of the surface. 
The slow migration of the Se vacancies \cite{TI_SbTe_exp,BiSe_vacancies_exp,BiSe_vacancies} results in an increased concentration of positively charged impurities close to the surface.
Thus, a time dependent BB was observed in ARPES experiments \cite{time_dep_ARPES,TI_SbTe_exp,SARPES_bite,BiSe_vacancies_exp,Cadopants_exp,Kdopants_exp1} where the bands at the surface were gradually bent downward.
In addition, it has been showed both experimentally \cite{Kdopants_exp1} and theoretically \cite{Kdopants,Kdopants2} that a potassium (K) layer deposited on the surface of \bise crystal triggers similar BB effects on the band structure as the Se vacancies.
Thus, the electron donor K adatoms can be also described by positively charged impurities close to the surface.

\subsection{Charge dopants on the surface of the  \bise crystal}

In our calculations we simulated the effect of the surface dopants by a planar capacitor situated on the surface of the \bise crystal, and charged by $\delta q_s$ per two-dimensional unit cell.
For example, the presence of the electron donor Se vacancies or K adatoms can be described by a positively charged capacitor.
On the other hand, the presence of the electron acceptor NO$_2$ molecules used in the ARPES experiments of Ref. \cite{SARPES_bite} can be modeled by a negatively charged capacitor.
This is a good assumption because the characteristic size of the probed samples are typically much larger than the lattice constant and the surface physics 
can be studied in terms of average physical quantities, such as the average concentration of the charged surface dopants.
In our model the charge of the capacitor corresponds to the average charge transfer between the dopants and the surface, while we neglect the inhomogeneities
on the atomic length scale.
Additionally,  since we are not confined by the supercell model of other first principle calculations \cite{Kdopants,Kdopants2}, 
we can now study the shift of the Dirac cone as a function of the doping in the entire concentration range.

The electric field of the charged capacitor is accounted for by means of a term in the Poisson equation, that is solved self-consistently within the SKKR code.
In the vacuum the electric field of the capacitor is canceled due to the attracted (or repelled) electrons from (to) the bulk reservoir. 
\begin{figure}[thb]
\centering
\includegraphics[scale=0.43]{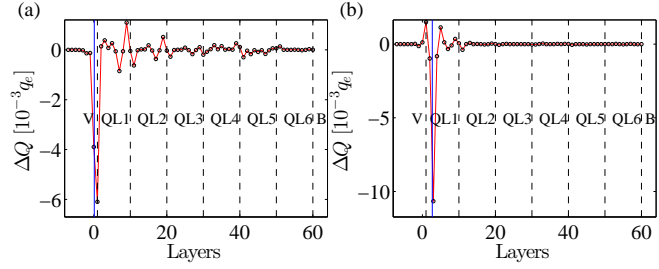}
\caption{Charge excess on the individual layers induced by a capacitor charged with \mbox{$\delta q=0.01q_e$}  per two-dimensional unit cell.
The interface region consists of six QLs, surrounded by the vacuum (V) from the left and by the bulk (B) from the right.
The solid vertical lines indicate the position of the capacitor in each figure.
} \label{fig:delta_charge}
\end{figure} 
Fig.~\ref{fig:delta_charge} shows the induced excess charge ($\Delta Q$) on the individual layers compared to the undoped system.
The results were obtained for a capacitor located close to the surface, and being charged to $\delta q=0.01q_e$, where $q_e\approx1.6\times10^{-19}$ is the elementary charge unit. 
In particular, Fig.~\ref{fig:delta_charge} shows the excess charge obtained for two different positions of the capacitor.
As expected, the charge of the capacitor is compensated by the accumulated electrons within the first two QLs, since the induced excess charge can be hosted only by the unsaturated SSs.
It can also be noticed that the induced effective electric field might generate further charge transfer between the layers in the low-lying QLs as well.
For example, in Fig.~\ref{fig:delta_charge}(a) one can clearly observe oscillations of $\Delta Q$ up to the $5$th QLs. 
Since the oscillations are centered around $\Delta Q=0$, the net charge of the corresponding QLs remains zero within the numerical precision of our calculations.
Our conclusions are consistent with previous theoretical works \cite{Kdopants,Kdopants2} predicting long range oscillations in the charge transfer.
However, our results indicate that these oscillations becomes suppressed for charge dopants located beneath the surface of the \bise crystal.
In the following we present our result obtained for the planar capacitor located at position corresponding to Fig.~\ref{fig:delta_charge}(a).

\begin{figure}[thb]
\centering
\includegraphics[scale=0.5]{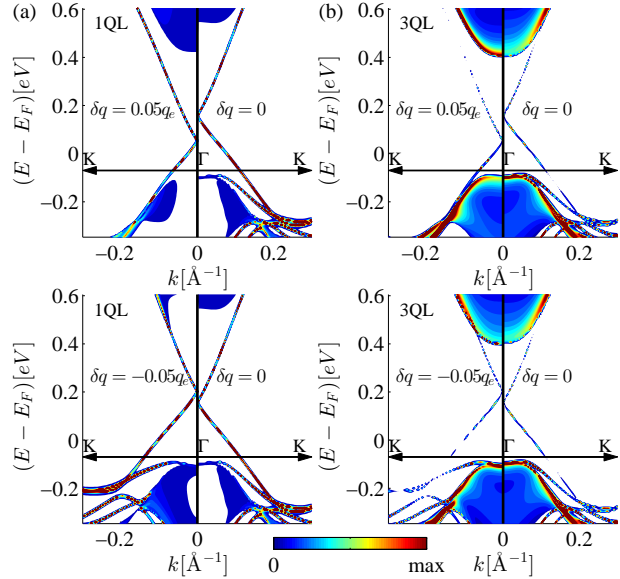}
\caption{Top panels: The BSF at positive surface doping compared to the clean system.
The BSF was calculated using Eq.~(\ref{eq:spectral}) and summed over the atomic layers in the (a) 1QL and (b) 3QL.
Bottom panels: Similar to the top panels but for negative surface doping.
The projection of the bulk bands onto the two-dimensional BZ is shown by colored areas, while the narrow lines correspond to the bands with low dispersion in the $z$ direction, including the SSs.
The BSF was calculated at complex energies with small imaginary part of $\sim0.7$ meV.
} \label{fig:spectral_surfdoping}
\end{figure}

Figure \ref{fig:spectral_surfdoping}. compares the band structures of the doped and clean systems.
For positively charged capacitor the accumulation of the electrons close to the surface is energetically favorable.
The states for these accumulated electrons are provided by the downward shift of the unsaturated SS bands [see Figs.~\ref{fig:spectral_surfdoping}.(a) and (b)].
Moreover, the BB of the bulk bands can also be tuned by a surface doping.
For a given concentration of the surface dopants, for example, the position of the conduction band at the surface can be shifted back to that in the bulk. 
For a negatively charged capacitor one can follow analogous reasoning.
In this case the bands are shifted upward as shown in Fig.~\ref{fig:spectral_surfdoping}(c) and (d).

In previous numerical studies the effective electric field responsible for the BB was controlled by a charge transfer originating from a sheet of adatoms located above the \bise surface \cite{Kdopants,Kdopants2,Bibilayer,spatial_Dirac_shift}.
In these calculations additional SSs were identified in the band structure being attributed to the presence of the given adatoms.
In the case of potassium for example, even quantum well states have been found within the bulk band gap where the Dirac cone is situated \cite{Kdopants}.
Moreover, due to a conventional insulator interface on the top of \bise, the protected SSs can even be spatially shifted towards the bulk of the \bise crystal \cite{spatial_Dirac_shift}.
Since we employ a different approach to model the BB effect, these artifacts are entirely absent from our results.
As a result, our method, enables us to study the evolution of the Dirac cone as a function of the surface doping in the entire concentration range, especially in the 
low concentration limit 
which is hardly accessible for approaches based on a supercell method due to numerical limitations.

The energy shift of the Dirac cone was successfully demonstrated by ARPES measurements as well \cite{SARPES_bite,Kdopants_exp1}.
As it is shown in Fig.~\ref{fig:spectral_surfdoping}, the surface bands are shifted upward or downward equally in each QL.
This behavior is in line with the localized nature of the surface states showing no momentum dispersion along the direction perpendicular to the surface.
On the other hand, the bend of the bulk bands varies continuously with the distance measured from the surface.
Indeed, in Fig.~\ref{fig:comparison}(a)-(d) one can observe a gradual decrease of the conduction band minimum in the successive QLs.
The layer resolved BSF (not shown in the manuscript) also confirms that the energy shift of the conduction band minimum varies smoothly from layer-to-layer.

We now examine the relation between the energy shift of the Dirac cone and the concentration of the charge dopants.
In case the Fermi energy is located inside the bulk band gap the excess charge induced by the surface doping can be hosted only by the unsaturated SSs, forming a Dirac cone at the center of the two-dimensional BZ.
One can then expect, that the energy shift of the Dirac cone is in correspondence with the increment or reduction of the occupied electron states required to host the induced excess charge.
The total charge per two-dimensional unit cell ($\Omega$) corresponding to the electron states on the Dirac cone between energies $E_1$ and $E_2$ can be calculated as
\begin{equation}
 \delta q_D = q_e\int\limits_{E_1}^{E_2}\rho_D(E) dE\;, \label{eq:q_dirac}
\end{equation}
where
\begin{equation}
 \rho_D(E) = \frac{\Omega}{(2\pi)^2}\oint\limits_{\Gamma_E}\frac{dk_{\parallel}}{\hbar v_g(\mathbf{k})}
\end{equation}
is the density of states and $\hbar v_g(\mathbf{k})=|{\rm grad}_{\mathbf{k}}E_{\pm}(\mathbf{k})|$ is the group velocity. 
The integral path is taken over the constant energy contour $\Gamma_E$ of the SS spectrum (\ref{eq:TI_spektrum}) at energy $E$.
\begin{figure}[thb]
\centering
\includegraphics[scale=0.5]{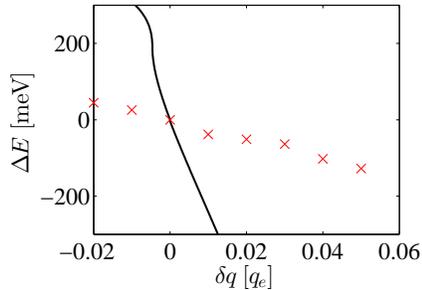}
\caption{Energy shift of the Dirac cone predicted by Eq.~(\ref{eq:q_dirac}) as a function of the surface doping per a two-dimensional unit cell (solid line).
Red crosses represent the results obtained by the SKKR calculations.
} \label{fig:Q_E}
\end{figure}
Figure \ref{fig:Q_E}. shows the calculated energy shift of the Dirac cone as a function of the deposited excess charge.
Surprisingly, we have found a significant difference between the predictions of the effective model and the SKKR results.
The effective model highly overestimates the energy shift of the Dirac cone compared to the results obtained by the SKKR model.
One can explain this inconsistency by the following arguments.
Within the effective model we assumed that the surface bands are shifted without any notable changes of the dispersion.
However, this approximation is valid only at energies close to the Dirac point.
Indeed, in Fig.~\ref{fig:spectral_surfdoping} the shape of the Dirac cones are similar in the doped and undoped cases.
One can, however, observe remarkable changes in the surface band structure around the crossing points with the other bands when surface dopants are present in the system (see Fig.~\ref{fig:spectral_surfdoping}).
Consequently, the charge density distribution on these parts of the band structure undergo to a significant change as well.
Thus, the low-energy segments of the unsaturated surface bands also play an important role in the screening of the surface dopants.

Comparing our results to other DFT based calculations \cite{Kdopants,Kdopants2}, generally we found a good qualitative agreement. The quantitative 
deviations can be related to the differences in the applied physical models.
While most of the DFT calculations use a supercell approach, our method relies on using semi-infinite bulk and vacuum regions.
Secondly, in Ref. \cite{Kdopants} the charge transfer between the deposited potassium atomic layer (K) and the \bise surface was controlled by the 
K-\bise distance instead of the K concentration on the surface.
In this work we found the position of the Dirac point to be more sensitive to the amount of charge transfer, resulting in a Dirac cone shift-charge transfer 
relation that is closer to the prediction of the effective model.
Our results, on the other hand, indicate that the energy shift of the Dirac cone cannot be estimated within the effective model given by 
Hamiltonian (\ref{eq:TI_FU_ham}), since a significant portion of the induced excess charge is hosted by the low-energy segments of the surface bands.
Thus, the position of the Dirac point is very much influenced by the treatment of the electrostatic potential used in the specific surface calculations.

\subsection{Spatially distributed charge dopants}

\begin{figure}[thb]
\centering
\includegraphics[scale=0.43]{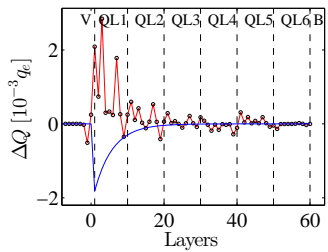}
\caption{Charge excess on the individual layers induced by dopants of total charge \mbox{$\delta q=-0.01q_e$} and being spatially distributed in the layers close to the surface.
The smooth blue line describes the concentration of the charge dopants as a function of the distance measured from the surface calculated by Eq. (\ref{eq:spatial_dopants}) using $\xi\approx5$ \AA.
The interface region consists of six QLs, surrounded by the vacuum (V) from the left and by the bulk (B) from the right.
} \label{fig:delta_charge_spatial}
\end{figure}

Besides a planar capacitor we also considered a scenario of spatially distributed charge dopants below the surface, that might be closer to the realistic case of an exposed surface.
One can expect that the concentration of the surface dopants vanish exponentially with the distance measured from the surface of the crystal.
Thus, in our calculations we described the concentration of the charged impurities by an exponential function
\begin{equation}
 \delta q_{\xi}(z) = \frac{\delta q}{\xi}{\rm Exp}\left(-\frac{z}{\xi}\right). \label{eq:spatial_dopants}
\end{equation}
The parameter $\xi$ describes the penetration depth of the dopants into the crystal.
The case of the planar capacitor can be recovered by the limit $\xi\rightarrow 0$.
In this section we discuss the results of our SKKR calculations with finite $\xi$ parameters.
In particular the layer dependent excess charge in Fig. \ref{fig:delta_charge_spatial} was calculated for charge dopants concentrated into the first two QLs below the surface.
As one can see in Fig. \ref{fig:delta_charge_spatial}, the oscillations in the excess charge decay on a length scale of $4-6$ QLs, which is longer that we have seen for a planar capacitor located inside the crystal.
Our calculations also indicate that the decaying length of the oscillations increases with the parameter $\xi$, which is consistent with our expectations.

\begin{figure}[thb]
\centering
\includegraphics[scale=0.5]{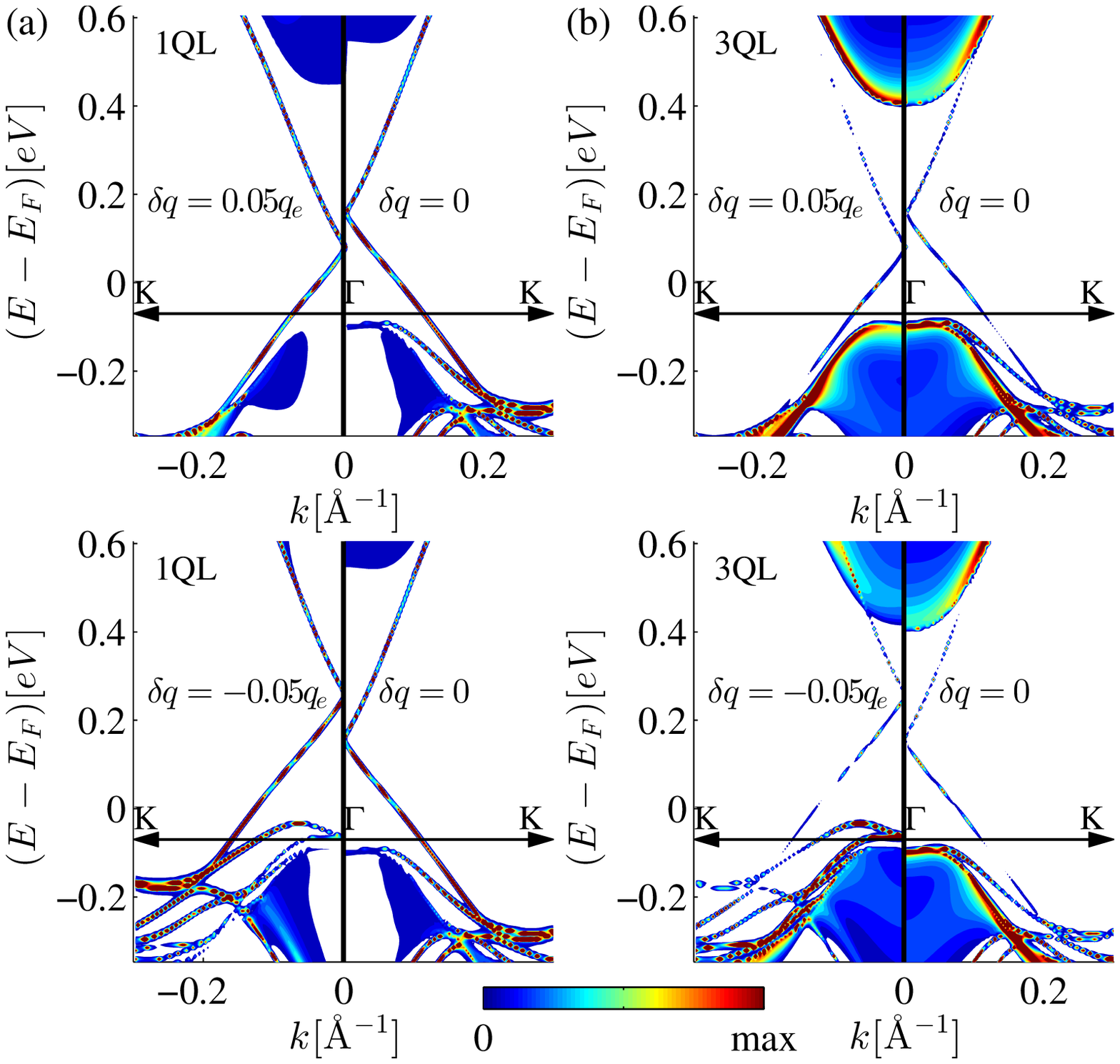}
\caption{Top panels: The BSF at positive spatial doping compared to the clean system.
The BSF was calculated using Eq.~(\ref{eq:spectral}) and summed over the atomic layers of (a) the 1QL and (b) 3QL.
Bottom panels: Similar to the top panels but for negative spatial doping.
The projection of the bulk bands onto the two-dimensional BZ is shown by a colored areas, while the narrow lines correspond to the bands with low dispersion in the $z$ direction, including the SSs.
The BSF was calculated at complex energies with small imaginary part of $\sim0.7$ meV, and $\xi\approx5$ \AA.
} \label{fig:spectral_surfdoping_spatial}
\end{figure}

Figure \ref{fig:spectral_surfdoping_spatial}. compares the band structures of the spatially doped and clean systems.
We found very similar results compared to the case when the charge dopants were modeled by a planar capacitor on the surface (see Fig. \ref{fig:spectral_surfdoping}.).
The downward shift of the Dirac cone induced by positive charge dopants [see Fig. \ref{fig:spectral_surfdoping_spatial}.(a) and (b)] is smaller, but very close to that induced by the planar capacitor used in the previous section [see Fig. \ref{fig:spectral_surfdoping}.(a) and (b)].
For negative charge dopants, on the other hand, the upward shift of the Dirac cone is about the twice of the shift induced by the planar capacitor.
However, this energy shift is still much less than the predicted value by the effective continuous model.
Thus, the qualitative conclusions made in the previous section are also applicable for the case of spatially distributed charge dopants.


\section{Summary}
\label{sec:summary}

In summary, we have calculated the band structure of \bise topological insulator by using the SKKR method.
In order to examine the effect of the charged impurities on the properties of the \bise surface, we also calculated the surface band structure in the presence of a charged planar capacitor situated close to the surface and for spatially distributed dopants under the surface.
We have found, that for a Fermi energy located in the bulk band gap the induced excess charge is hosted by the unsaturated SSs.
Thus, the charge of the surface dopants is also screened within the first few QLs below the surface.
In addition, due to the excess charge and the inhomogeneous charge distribution close to the surface, the bulk bands undergo a BB effect even in the pristine \bise crystal.
Consequently, the bulk bands becomes bent in the atomic layers close to the surface, but one can already recover the properties of the bulk crystal starting from the third QL.
Our results also indicate, that the BB profile can be tuned via the deposited surface dopants.

In contrast to the bulk bands, the Dirac cone (being formed by the SS bands inside the bulk band gap) becomes shifted in energy due to the deposited surface dopants.
The magnitude and the direction of this energy shift depends on the concentration and on the sign of the deposited dopants.
However, this effect cannot be described within the scope of the effective continuous model of the SSs.
Our self-consistent numerical results showed that besides the Dirac cone the low-energy segments of the surface bands also play an important role in the electrostatic properties of the surface.
We also found that in agreement with recent theoretical studies \cite{Kdopants,Kdopants2}, the charged impurities on the surface induce oscillations in the charge density extending deep into the crystal.
However, our results indicate that in the case when the dopants are located beneath the surface, these oscillations became highly suppressed.

In order to check experimentally our findings, one needs to independently measure the doping concentration on the surface and the energy shift of the Dirac cone.
We believe that the combination of the scanning tunneling microscope and ARPES techniques can serve this purpose.
Moreover, the surface doping of the 2GTIs can be used to cancel the BB effect on the surface, which is essential to take an advantage of the protected SSs in transport devices.
Finally, the possibility of tuning the position of the Dirac cone might also be of great importance for future experimental applications of these materials.

\section*{Acknowledgments}

We acknowledge the support from the Hungarian Scientific Research Fund No. OTKA K115575, K115632 and K108676.


\appendix

\section{The bulk band structure of \bise compound} \label{subsec:bulk}

In this section we determine the band structure of the bulk \bise crystal. 
To this end we calculate the layer resolved BSF given by Eq.~(\ref{eq:spectral}) for the bulk system at complex energies with a small imaginary part, and summed the contributions of the individual atomic and empty layers within one QL. 
\begin{figure}[thb]
\centering
\includegraphics[scale=0.6]{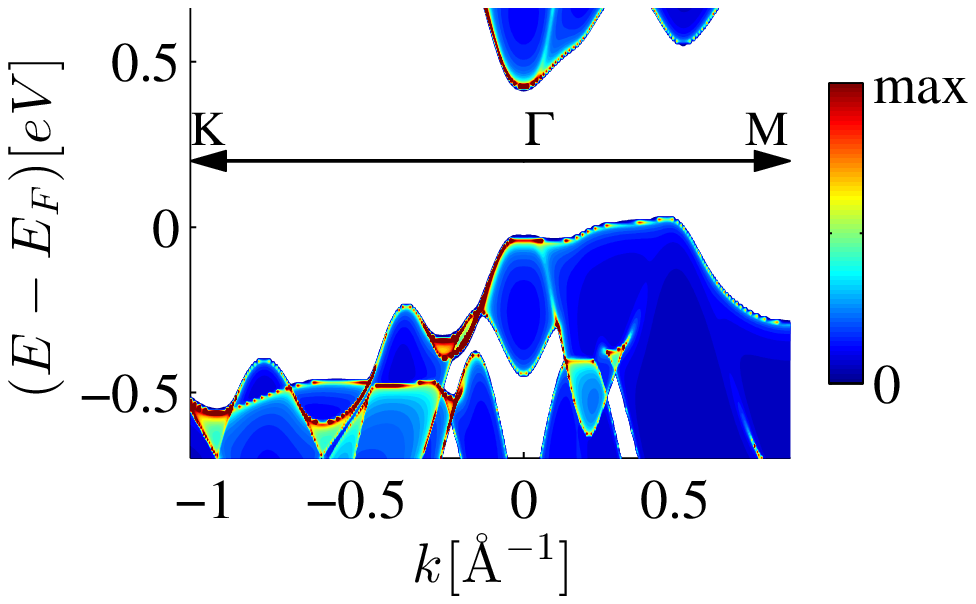}
\caption{Projection of the BSF onto the two-dimensional BZ summed over the layers of one QL inside the bulk plotted along the $K\Gamma M$ cross section of the two-dimensional BZ.
The blue and red colors represents areas of low and high density of states, respectively.
The Fermi level is at the top of the valence band. 
The BSF was calculated at complex energies with small imaginary part of $\sim0.7$ meV.
} \label{fig:spectral_bulk}
\end{figure}
The result of this procedure can be interpreted as the projection of the bulk band structure onto the two-dimensional BZ corresponding to the crystal intersection formed perpendicularly to the $z$ direction. (For details see the main text.)
The calculated BSF along the $K\Gamma M$ cross section of the BZ is shown in Fig.\ref{fig:spectral_bulk}.
The high density areas (forming narrow lines close to the extremal points of the band structure) indicate electron states of low dispersion along the $z$ direction.
The Fermi level is at the top of the valence band. 

It should be noted that the Fermi level is sensitive to the presence of impurities, even at small concentration.
Several ARPES measurements indicated that due to the bulk dopants the Fermi energy was tuned into the bulk conduction band.\cite{bise_arpes}
According to our calculations, the direct bulk band gap at the $\Gamma$ point is $\sim447$ meV.
Our results are in good agreement with other theoretical calculations.\cite{TI_Zhang,BiSe_lattice,gap_theo2}


\end{document}